\RequirePackage{lineno}
\documentclass[twocolumn,nofootinbib,floatfix,superscriptaddress]{revtex4-1}

\usepackage{amsmath,amssymb}
\usepackage{cancel}
\usepackage{graphicx}
\usepackage{epsfig}
\usepackage{amsmath,amsfonts,amssymb}
\usepackage{multirow}
\usepackage{colortbl}
\usepackage{hhline}
\usepackage{booktabs}
\usepackage{bbm}
\usepackage{bm}
\usepackage{tabulary}
\usepackage{pifont}
\usepackage{bbold}
\usepackage{pifont}
\usepackage{soul}

\newcolumntype{K}[1]{>{\centering\arraybackslash}m{#1}}
\begin{document}

\title{$(g-2)_\mu$ in the 2HDM and slightly beyond -- an updated view}

\author{P. M. Ferreira}
\affiliation{Instituto Superior de Engenharia de Lisboa, Instituto Polit\'ecnico de Lisboa 1959-007 Lisboa, Portugal}
\affiliation{Centro de F\'{\i}sica Te\'orica e Computacional, Faculdade de Ciências,
Universidade de Lisboa, Campo Grande, Edif\'{\i}cio C8 1749-016 Lisboa, Portugal}
\author{B. L. Gon\c{c}alves}
\affiliation{Departamento de F\'{\i}sica and CFTP, Instituto Superior T\'ecnico, Universidade de Lisboa, Lisboa, Portugal}
\affiliation{Centro de F\'{\i}sica Te\'orica e Computacional, Faculdade de Ciências,
Universidade de Lisboa, Campo Grande, Edif\'{\i}cio C8 1749-016 Lisboa, Portugal}
\author{F. R. Joaquim}
\affiliation{Departamento de F\'{\i}sica and CFTP, Instituto Superior T\'ecnico, Universidade de Lisboa, Lisboa, Portugal}
\author{Marc Sher}
\affiliation{High Energy Theory Group, William \& Mary, Williamsburg,
VA 23187, USA}


\begin{abstract}
The recent measurement of the muon $g - 2$ anomaly continues to defy a Standard Model explanation, but can be accommodated within the framework of two Higgs doublet models, although the pseudoscalar mass must be fairly
light. If one further includes extra fermion  content in the form of a generation of vector-like leptons, the allowed parameter range that explains the anomaly is even further extended, and clashes with $B$-decay constraints may be avoided. We will show how the muon magnetic moment anomaly can be fit within these models, under the assumption that the vector-like leptons do not mix with the muon. We update previous analyses and include all theoretical and experimental constraints, including searches for extra scalars. It is shown that the inclusion of vector-like fermions allows the lepton-specific and muon-specific models to perform much better in fitting the muon's $g - 2$. However, these fits do require the Yukawa coupling between the Higgs and the vector-like leptons to be large, causing potential problems with perturbativity and unitarity, and thus models in which the vector-like leptons mix with the muon may be preferred.
\end{abstract}

\maketitle

\section{Introduction}

Recently, the Muon g-2 Collaboration at Fermilab reported new results~\cite{results} from Run 1 of their experiment measuring the anomalous magnetic moment of the muon $a_\mu$. Prior to this announcement, the discrepancy between the experimental measurement $a_\mu^{\rm exp}$~\cite{Bennett:2006fi}  and the Standard Model (SM) theoretical prediction $a_\mu^{\rm SM}$ ~\cite{Blum:2018mom,Keshavarzi:2018mgv,Davier:2019can,Aoyama:2020ynm} was
\begin{equation}
\Delta a_\mu^{\rm exp}=a_\mu^{\rm exp}-a_\mu^{\rm SM}= (279 \pm 76) \times 10^{-11}\quad (3.7 \sigma),
\label{eq:old}
\end{equation}
while the new combined result is~\cite{results}
\begin{equation}
\Delta a_\mu^{\rm exp} = (251 \pm 59) \times 10^{-11}\qquad (4.2\sigma).
\label{eq:new}
\end{equation}
There are hundreds of papers with new-physics explanations for the $(g-2)_\mu$ anomaly such as supersymmetric models, left-right symmetric models, scotogenic models, 331 models, $L_\mu - L_\tau$ models, seesaw models, the Zee-Babu model as well as two-Higgs doublet models (2HDMs); an extensive review can be found in Ref. \cite{Lindner:2016bgg}. In this paper, we will focus on 2HDMs (for a review, see Ref.~\cite{Branco:2011iw}), and discuss the implications of the new result from the Muon g-2 Collaboration.

In the 2HDM it is possible to ensure that tree-level flavor changing neutral currents mediated by scalars do not exist, by imposing a discrete $Z_2$ symmetry on the model. There are four such versions of the 2HDM,  referred to as the type-I, type-II, type-X (sometimes called lepton-specific) and type-Y (sometimes called flipped) models. In the type-II and type-X models, the coupling of the muon to the heavy Higgs bosons is enhanced by a factor of $\tan\beta$, which is the ratio of the two vacuum expectation values, but in type-II models, the $Q=-1/3$ quark couplings also get the same enhancement, leading to possible problems with radiative $B$-meson decays.  In fact, explaining the $(g-2)_\mu$ anomaly without affecting $B$-decays is one of the motivations for studies of the type-X model. On the other hand, in the type-I and type-Y models, the couplings of the muon to heavy Higgs bosons are suppressed by $\tan\beta$ and, thus, these models are not favored for explaining the $(g-2)_\mu$ discrepancy. For simplicity, we will only consider models without tree-level flavor-changing neutral currents and with CP conservation in the Higgs sector.

One of the earlier $(g-2)_\mu$ studies in 2HDMs following the discovery of the light Higgs boson was the work of Broggio et al. \cite{Broggio:2014mna}.    They restricted their analysis to the ``alignment limit" ($\cos(\beta-\alpha )= 0$) in which the tree-level couplings of the lighest 2HDM scalar are identical to those of the SM Higgs. As noted above, they found that only the type-II and type-X models could account for the $(g-2)_\mu$ discrepancy. In both models, a very light pseudoscalar Higgs mass ($m_A$) is required, typically below $100$ GeV,  as is a relatively large $\tan\beta$, typically greater than $60$.  
Another analysis of the type-X 2HDM with low-mass pseudoscalars is that of~\cite{Wang:2014sda}, where it was shown that $m_A$ could be as low as 10 GeV. With values for $m_A$ as in these references, unitarity and electroweak precision constraints then force the 
charged Higgs mass ($m_{H^+}$) to be less than about $200$ GeV. This is a problem within the type-II model, since radiative $B$ decays, $\Delta m_{B_s}$ and the hadronic $Z\rightarrow \bar{b}b$ branching ratio force $m_{H^+} \gtrsim 600$~GeV~\cite{Arbey:2017gmh}. As a result, the type-X model is favored. A subsequent analysis in Ref. \cite{Abe:2015oca} focusing explicitly on the type-X model considered all experimental constraints. It was noted that bounds from leptonic $\tau$ decay restricted the parameter-space further and that the discrepancy in $(g - 2)_\mu$  could be explained at the $2 \sigma$ level with $m_A$ between $10$ and $30$ GeV, $m_{H^+}$ between $200$ and $350$ GeV and $\tan\beta$ between $30$ and $50$. Shortly thereafter, Ref.~\cite{Chun:2016hzs} included more recent data from lepton universality tests and found bounds that were somewhat weaker, but in general agreement with \cite{Abe:2015oca}. In the same token, studies have been performed in Refs.~\cite{Chun:2015hsa,Ilisie:2015tra}, and a more recent work including limits from Higgs decays to $AA$~\cite{Wang:2018hnw} also found similar restrictions on the parameter-space.

Another attempt to explain the $(g-2)_\mu$ discrepancy involved adding vector-like leptons (VLL). It was shown~\cite{Dermisek:2013gta,Falkowski:2013jya} that this alternative works if the VLLs mix with the muon. However, in the SM this not only alters the Higgs dimuon branching ratio but also affects excessively the diphoton branching ratio and is thus phenomenologically unacceptable. Recently, the addition of VLLs to the type-II and type-X models was considered~\cite{Frank:2020smf} and it was shown that the parameter-space can be significantly expanded even without mixing the VLLs with muons, since the VLLs contribute to $\Delta a_\mu$ in two-loop Barr-Zee diagrams. In this case, the parameter-space of the type-X model is substantially widened and the type-II model is not completely excluded. Even more recently, Dermisek, Hermanek and McGinnis~\cite{Dermisek:2020cod} explored the type-II model with VLLs, but now allowed mixing with the muon. Here, the effects on the muon coupling to the weak vector and scalar bosons must be considered. They showed that the extra Higgs bosons and VLLs could be extremely heavy and still explain the $(g-2)_\mu$ discrepancy without major effects on the dimuon SM Higgs decay. A more detailed and comprehensive paper by Dermisek et al. has just appeared~\cite{Dermisek:2021ajd}.
None of the above papers deal with the smaller discrepancy for the $(g-2)$ of the electron. A model explaining both discrepancies was proposed by Chun and Mondal~\cite{Chun:2020uzw} in which VLLs mix with the muon and the electron. It turns out that also in such case a light $(<100\,{\rm GeV})$ pseudoscalar and large $\tan\beta$ are needed.

The belief that only the standard four 2HDMs can avoid flavor-changing neutral currents at tree level was shown to be incorrect by Abe, Sato and Yagyu \cite{Abe:2017jqo}.  They proposed a muon-specific ($\mu{\rm Spec}$) model in which one  doublet couples to the muon and the other couples to all other fermions.  This was implemented with one of the usual $Z_2$ symmetries, complemented with a muon number conservation symmetry (which is a global $U(1)$, although they referred to it as an overall $Z_4$ symmetry). This symmetry cannot be implemented in the quark sector since it eliminates CKM mixing, but neutrino mixing can easily be produced in a heavy Majorana sector. They studied $(g-2)_\mu$ in the model and found that the discrepancy could be solved with very large $\tan\beta$ values of approximately $1000$ (but they also showed that perturbation theory was still valid). A study of the model by PF and MS~\cite{Ferreira:2020ukv} showed that this was fine-tuned and, leaving aside any explanation of the $(g-2)_\mu$ discrepancy, studied other properties of the model, including substantial effects on the dimuon decay of the Higgs. 

In the above, we have only considered 2HDMs in which a symmetry eliminates tree level flavor changing neutral currents.    An alternate approach is to simply assume that the Yukawa coupling matrices of the two doublets are proportional.   This is the Aligned 2HDM \cite{Pich:2009sp,Jung:2010ik} (A2HDM).    The conventional 2HDMs are special cases of the A2HDM.
Early treatments of $(g-2)$ in this model were discussed in Refs. \cite{Ilisie:2015tra,Han:2015yys}.    Since the parameter-space is larger one has an expanded range of allowed pseudoscalar masses (although the pseudoscalar mass must still be relatively light) and can accommodate smaller $\tan\beta$. Notice that 
$\tan\beta$ is defined in the A2HDM in the  usual manner (the ratio of doublet vevs), but the scalar-fermion couplings do not depend on it in the same way as in the flavour conserving 2HDMs; while in the latter models the up, down  and lepton yukawa couplings have correlated $\tan\beta$ dependencies, in the A2HDM those couplings are essentially independent, which clearly increases the allowed parameter space.
A much more detailed two-loop computation in the A2HDM was in Ref. \cite{Cherchiglia:2016eui}, and a complete phenomenological analysis, focusing on the effects of a light pseudoscalar, was in Ref. \cite{Cherchiglia:2017uwv}.   The papers of \cite{Han:2015yys,Cherchiglia:2017uwv} also studied the particularly interesting decay of the Higgs into $AA$, which is important if the pseudoscalar is below $62$ GeV.     In this paper, we will focus on the models in which a symmetry forces tree level FCNC to vanish but will mention the A2HDM in the conclusions.

Finally, one must take into account that there are many alternative attempts to explain $(g-2)_\mu$, even 
in the context of 2HDMs. For instance, the model of~\cite{Han:2018znu} has a lepton-specific inert doublet with $\mu$ Yukawa 
couplings having an opposite sign to those of $e$ and $\tau$; models with tree-level FCNC were considered
in this context in~\cite{Crivellin:2015hha,Crivellin:2019dun}; in~\cite{Arnan:2019uhr} the impact of a fourth generation 
of vector-like fermions and an extra scalar was studied; contributions from additional 
gauge scalar singlets were investigated in~\cite{vonBuddenbrock:2016rmr,Sabatta:2019nfg}; an extremely light (sub-GeV) 
scalar was considered in~\cite{Jana:2020pxx}; a 2HDM study of $(g-2)_\mu$ considering the effects of
warped space was undertaken in~\cite{Megias:2017dzd}; and a 2HDM complemented with a fourth generation
of fermions and a $Z^\prime$ gauge boson from an extra $U(1)_X$ symmetry was considered in~\cite{Chen:2019wbk}. 
As for non-2HDM explanations, one has, for instance, an addition of a leptophilic scalar to the SM in~\cite{Chen:2021rnl}; UV complete models with a $L_\mu-L_\tau$ symmetry were considered in~\cite{Crivellin:2018qmi}; a simultaneous study of dark matter and $(g-2)_\mu$ with extra scalars and vector-like fermions was undertaken in \cite{Jana:2020joi}; a generic analysis of radiative leptonic-mass generation and its impact on $(g-2)_\mu$ can be found in~\cite{Yin:2021yqy};
and the interplay between new physics contributions and the SM effective field theory for $(g-2)_\mu$ is discussed 
in~\cite{Fajfer:2021cxa}. On a different strand, the experimental implications of the muon anomalous magnetic moment at a future muon collider were studied in~\cite{Capdevilla:2020qel,Buttazzo:2020eyl,Yin:2020afe}.

All of the above analyses relied on the old experimental result~\cite{Bennett:2006fi} and one would expect the new result from the Muon g-2 Collaboration~\cite{results} to have a small effect on those studies. In this paper, we update some of the 2HDM results to include the new combined experimental value for $\Delta a_\mu^{\rm exp}$ given in \eqref{eq:new}. After discussing the impact of the new measurement in the context of the type-X and type-II models, we will focus on the 2HDM extended with VLLs which do not mix with the muon. Adding such mixing increases the number of parameters, and many of the relevant formulae are in the recent work of Dermisek et al.~\cite{Dermisek:2021ajd}. In addition, we also look at the $\mu{\rm Spec}$ model, which can suffer substantial changes in the Higgs dimuon decay, and analyse the $(g-2)_\mu$ discrepancy when we add VLLs. 
 
\section{Type-II, type-X and $\bm{\mu{\rm Spec}}$ 2HDMs}

In the 2HDM the scalar-fermion Yukawa interactions can be expressed in the compact form:
\begin{align}
    \mathcal{L}=&\dfrac{\sqrt{2}}{v}H^+ \left\{ \bar{u} \left[\xi_u M_u^\dag V P_L-\xi_d  V M_d P_R\right]d-\xi_\ell \bar{\nu} M_\ell P_R \ell\right\}\nonumber \\
   &-\dfrac{1}{v} \sum_{k,f} y_f^k S_k^0 \bar{f}M_f P_R f+{\rm H.c.}\,,
   \label{eq:LY}
\end{align}
where $S_k^0=\{h,H,A\}$ are the neutral-scalar mass eigenstates ($h$ is the SM Higgs with mass $m_h=125.1$~GeV~\cite{Zyla:2020zbs}), and $f=u,d,\ell$ denote any SM charged-fermion type with mass matrix $M_f$. Following the notation of Ref.~\cite{Pich:2009sp}, the couplings $y_{f,\ell}^k$ are 
\begin{equation}
\begin{aligned}
y_{d,\ell}^k&=\mathcal{R}_{k1}+(\mathcal{R}_{k2}+i\,\mathcal{R}_{k3})\,\xi_{d,\ell}\,,\\
y_{u}^k&=\mathcal{R}_{k1}+(\mathcal{R}_{k2}-i\,\mathcal{R}_{k3})\,\xi_{u}^\ast\,,
   \label{eq:Ycoup}
\end{aligned}
\end{equation}
where $\mathcal{R}$ is the orthogonal matrix which relates the scalar weak states with the mass eigenstates $S_k^0$. For a CP-conserving potential, $A$ does not mix with $h$ and $H$ implying $\mathcal{R}_{3j}=\mathcal{R}_{j3}=0$ for $j\neq 3$ and $\mathcal{R}_{33}=1$. On the other hand, $h-H$ mixing is determined by $\mathcal{R}_{11}=-\mathcal{R}_{22}=\sin(\beta-\alpha)$ and $\mathcal{R}_{12}=\mathcal{R}_{21}=\cos(\beta-\alpha)$, with $\beta - \alpha = \pi/2$ in the alignment limit. For the type-II, type-X and $\mu{\rm Spec}$ 2HDM, the $\xi_f$ couplings are
\begin{equation}
\begin{aligned}
{\rm Type \,II:}&\; \xi_{d,\ell}=-\tan\beta\;,\;\xi_{u}=\cot\beta\,,\\
{\rm Type \,X:}&\; \xi_{\ell}=-\tan\beta\;\;\;\,,\;\xi_{u,d}=\cot\beta\,,\\
\mu{\rm Spec:} &\; \xi_\mu=-\tan\beta\;\,\,\,,\;\xi_{u,d,\tau,e}=\cot\beta\,.
   \label{eq:xicoup}
\end{aligned}
\end{equation}
In the alignment limit, the tree-level couplings of the SM-like scalar state, $h$, are therefore
identical to those of the SM Higgs boson -- the agreement of observed Higgs production and decays to that of the SM makes this assumption quite reasonable.
\begin{figure}[t!]
	\centering
	\begin{tabular}{l}
		\includegraphics[width=0.48\textwidth]{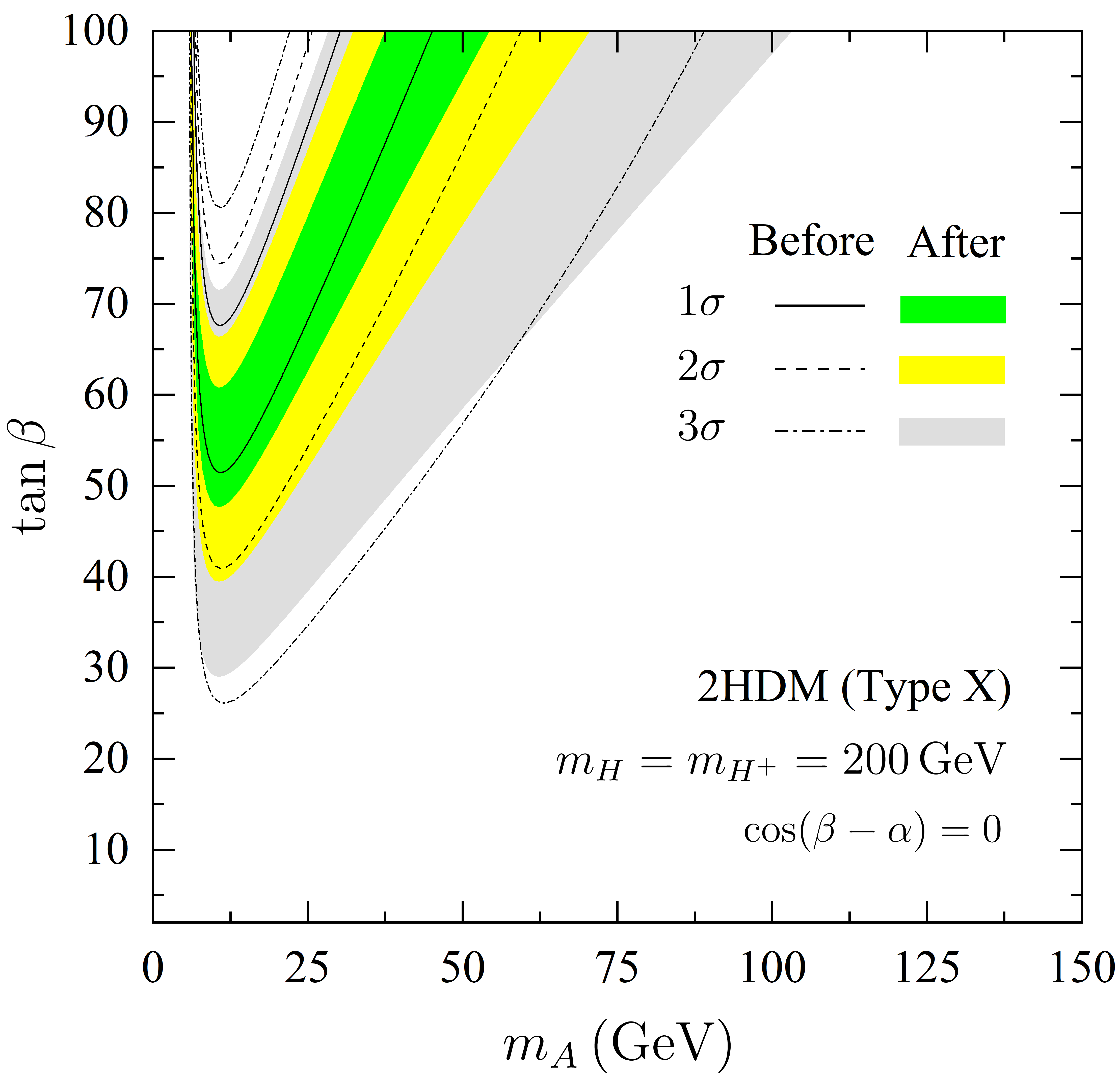}\\
	\includegraphics[width=0.48\textwidth]{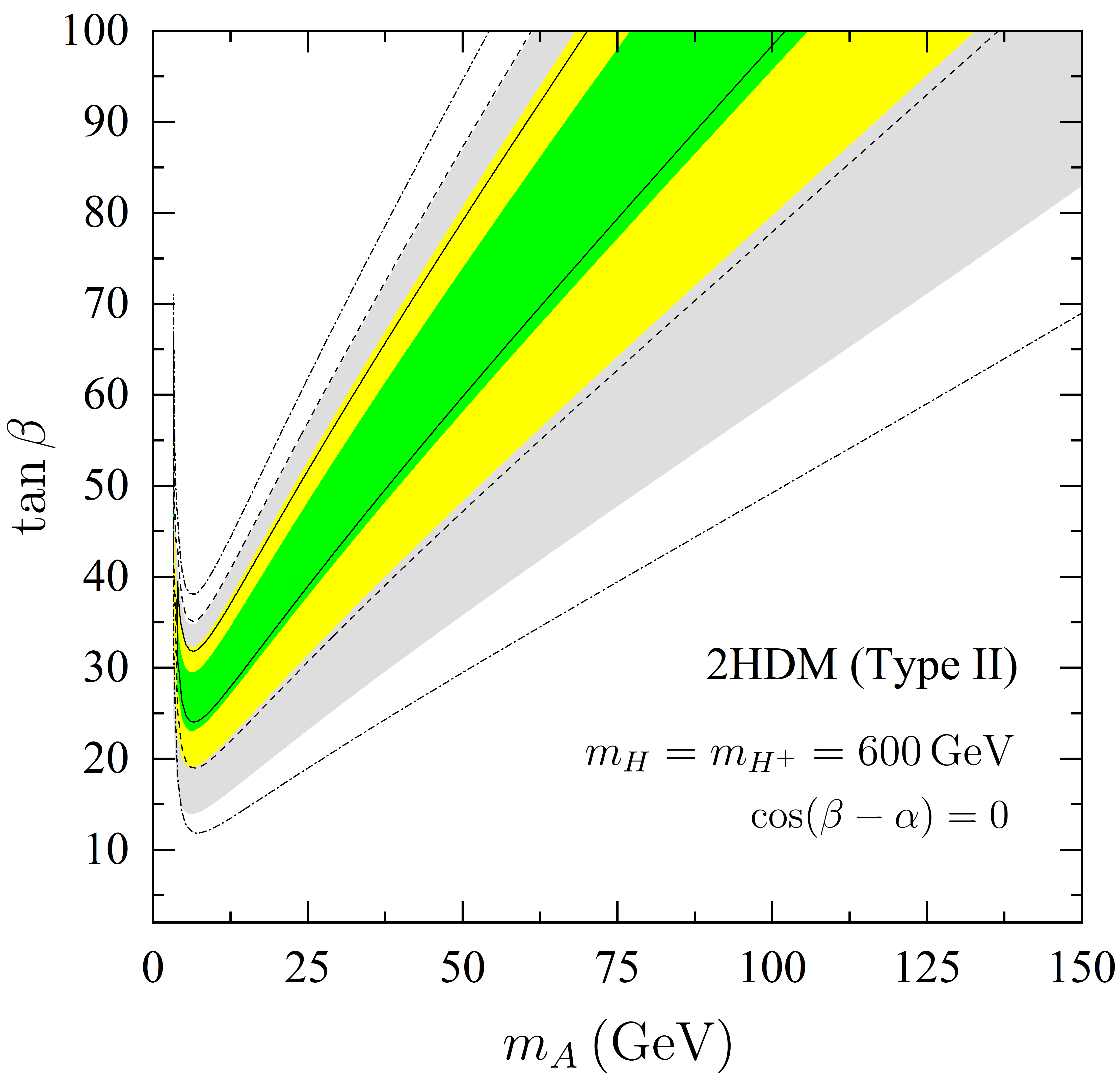} 
	\end{tabular}
\caption{$\Delta a_\mu$ allowed regions in the $(m_A,\tan\beta)$ plane for the type-X (top) and type-II (bottom) 2HDMs. The green, yellow and grey shaded regions were obtained taking the new $\Delta a_\mu$ interval given in Eq.~\eqref{eq:new} at 1,2 and $3\sigma$, respectively. For comparison, we also show the lines which delimit the same regions when the old result in Eq.~\eqref{eq:old} is considered.}
\label{fig1}
\end{figure} 

In the early post-Higgs discovery study of the $(g-2)_\mu$ anomaly of Ref.~\cite{Broggio:2014mna}, the type-II and type-X models were studied in the alignment limit.  The relevant free parameters in the model are $\tan\beta$, the pseudoscalar mass $m_A$, the charged Higgs mass $m_{H^+}$, the heavy scalar mass, $m_H$, and the $Z_2$ soft-breaking parameter $m^2_{12}$. Broggio at al. considered constraints on these parameters due to electroweak precision tests, vacuum stability and perturbativity. These  imposed bounds on the heavy Higgs masses, and they showed that for a pseudoscalar mass of less than $100$ GeV, the charged Higgs mass could not exceed $200$ GeV.   This was not substantially affected by the value of $\tan\beta$ or by deviations from the $\cos(\beta-\alpha)=0$ assumption. They then calculated the constraints from the $(g-2)_\mu$ results (they also included constraints from the $g-2$ of the electron, but these are very weak).     

We now revisit these findings 
in light of the new Muon g-2 Collaboration result considering all one-loop and two-loop Barr-Zee (BZ) contributions to $\Delta a_\mu$ as given in Ref.~\cite{Ilisie:2015tra}. In Fig.~\ref{fig1} we show the allowed regions in the $m_A$ and $\tan\beta$ plane corresponding to the $\Delta a_\mu$ intervals in Eq.~\eqref{eq:new} at the $1\sigma$, $2\sigma$ and $3\sigma$ levels (green, yellow and grey shaded regions) for the type-X (top panel) and type-II (bottom panel) cases. The solid, dashed and dash-dotted contours delimit the analogue regions when the old result in Eq.~\eqref{eq:new} is considered. As noted earlier, radiative $B$-decays favor $m_{H^+}>600$~GeV in the type-II model \cite{Arbey:2017gmh}. Thus, we take $m_{H^+}=600$~GeV as reference value in the bottom panel of Fig.~\ref{fig1}, while for type X a lighter $H^+$ is considered (changing $m_{H^+}$ does not have a significant impact on $\Delta a_\mu$). As pointed out in the Introduction, such small values of $m_A$ in the type-II model are in conflict with unitarity and electroweak precision constraints. On the other hand, type X is still allowed. The results show that, as expected, the impact of the new $(g-2)_\mu$ result is marginal. The $\mu$Spec 2HDM requires extreme fine-tuning and values of $\tan\beta$ of $O(1000)$ in order to accommodate the $(g-2)_\mu$ discrepancy, and we do not show a plot in this case. It is worth mentioning that for $\mu$Spec the BZ contributions to $\Delta a_\mu$ are not significant as they are for type II and type X since both the $b$-quark and $\tau$-lepton couplings are not $\tan\beta$ enhanced.

As a check, by keeping the fermion BZ diagrams only, as done by Broggio et al., we were able to reproduce their results for the type-II and type-X 2HDM in the alignment limit.  Fig. ~\ref{fig1} updates their work and includes additional contributions. Although Broggio et al. did include electroweak precision fits, other constraints arise from B-physics, Z-physics and $\tau$-physics. These are discussed in the A2HDM, which contains the type-II and type-X models as special cases, in Ref. \cite{Cherchiglia:2017uwv}. For the type-X model, the bounds in Fig. 1 of that work lead to an upper bound on $\tan\beta$ which rises from $30$ to $50$ as $m_A$ rises from $0$ to $20$ GeV and then levels off at a value between $60$ and $100$, depending on the mass of the heavier scalars.   This may cut off the upper part of the allowed parameter-space in Fig. ~\ref{fig1}~\footnote{After this paper was submitted for publication we became 
aware of~\cite{Athron:2021iuf}, where further updates to the 2HDM fit of $a_\mu$ are discussed.}. 

\section{2HDM with vector-like Leptons}

Let us now consider a 2HDM extension with vector-like leptons $\chi_{L,R}=(N\,L^-)_{L,R}^T\sim (\bm{2},-1/2)$ and $E_{L,R}\sim (\bm{1},-1)$, with  $N_{L,R}$ being neutral and $L^-_{L,R},E_{L,R}$ charged. If taken within the context of the muon-specific 2HDM, the VLLs will have no quantum numbers
under the underlying  muon number conservation  symmetry of that model. We choose to have no mixing between the VLLs and the usual 
leptons -- this can be achieved, for instance, by imposing an extra $Z_2$ symmetry on the model, under which all 
VLL fields have charge -1, and all other fields have charge +1.     
As noted in the introduction, the field content of the model is the same as in Refs. \cite{Frank:2020smf,Dermisek:2020cod,Dermisek:2021ajd,Chun:2020uzw}.   Refs. \cite{Dermisek:2020cod,Dermisek:2021ajd} do not have the extra $Z_2$ symmetry and thus allow mixing with the muon, while the other two references forbid such mixing.

The VLL Yukawa and mass terms are 
\begin{align}
    -\mathcal{L}_{\rm VLL} =& m_L \overline{\chi_{L}}\chi_R+m_E \overline{E_L}E_R
    +\lambda_L\overline{\chi_{R}}\Phi_1 E_L\nonumber \\
   &+\lambda_R\overline{\chi_{L}}\Phi_1 E_R +{\rm H.c.}\,,
   \label{eq:LYVLL}
\end{align}
which, after electroweak symmetry breaking, lead to the following heavy charged-lepton mass matrix
\begin{align}
    \mathcal{M}=\begin{pmatrix}
m_L& \lambda_R v \cos\beta/\sqrt{2}\\
\lambda_L^\ast v \cos\beta/\sqrt{2} &m_E
\end{pmatrix}\,,
   \label{eq:MVLL}
\end{align}
in the $(L^-\;E)_{L,R}^T$ basis, while for the new neutral-lepton mass $M_N=|m_L|$. The above matrix can be diagonalized as $U_L^\dag \mathcal{M} U_R={\rm diag}(M_1,M_2)$, where $M_{1,2}$ are the masses of the $L_{1,2}^-$ heavy charged-lepton physical states and $U_{L,R}$ are $2\times 2$ unitary matrices. As for the new charged-current interactions, we have
\begin{align}
    \!\!\!\!\mathcal{L}=\dfrac{g}{\sqrt{2}}\bar{N}\gamma^\mu\left[ (U_L)_{1a} P_L + (U_R)_{1a} P_R \right] L_a^-W_\mu^++{\rm H.c.},
   \label{eq:LCC}
\end{align}
where $g=e/\sin\theta_W$ is the SU(2)$_L$ gauge coupling. The interactions with the neutral and charged physical scalars are
\begin{align}
    -\mathcal{L}=&\dfrac{\sqrt{2}}{v}H^+ \bar{N} \left[\,\xi_a M_a (U_R)_{2a} P_R+\xi_N M_N (U_L)_{2a} P_L\right]L_a^-\nonumber \\
   &+\dfrac{1}{v} \sum_{k,a,b} M_a \,y_{ab}^k \overline{L^-_a} P_R L^-_b S_k^0+{\rm H.c.}\,,
   \label{eq:LSHVLL}
\end{align}
with $S_k^0=\{h,H,A\}$, $a,b=1,2$ and
\begin{equation}
\begin{aligned}
\xi_a&=\xi_\ell \frac{\mathcal{M}_{12}}{M_a}\;,\;
\xi_N=\xi_\ell \frac{\mathcal{M}_{21}^\ast}{M_N}\;,\; \xi_\ell=-\tan\beta\,,\\
y_{ab}^k&=(\mathcal{R}_{k1}-\xi_\ell \mathcal{R}_{k2})X_{ab}^++i\,\xi_\ell\,\mathcal{R}_{k3}X_{ab}^-\,.
\label{eq:xiVLL}
\end{aligned}
\end{equation}
For simplicity, we will restrict ourselves to the case of real $\mathcal{M}$, for which $U_{L,R}$ are orthogonal and parametrized by two mixing angles $\theta_{L,R}$ with $(U_{L,R})_{11}=(U_{L,R})_{22}=c_{L,R}$ and $(U_{L,R})_{12}=-(U_{L,R})_{21}=s_{L,R}$ (the notation $\cos\theta_{L,R}\equiv c_{L,R}$, $\sin\theta_{L,R}\equiv s_{L,R}$ has been used). Under this assumption, and taking into account that the only relevant VLL couplings are the flavor-conserving ones ($a=b$), we have:
\begin{equation}
\begin{aligned}
X_{aa}^+&=c_R^2s_L^2+c_L^2s_R^2-\dfrac{1}{2}s_{2L}s_{2R}\left(\dfrac{M_1}{M_2}\right)^{(-1)^a}\,,\\
X_{aa}^-&=\dfrac{(-1)^{a+1}}{2}(c_{2L}-c_{2R})\,,
\label{eq:xpm}
\end{aligned}
\end{equation}
with $c_{2L,2R}\equiv \cos(2\theta_{L,R})$ and  $s_{2L,2R}\equiv \sin(2\theta_{L,R})$. 
The rotation angles also allow us to express the couplings $\lambda_{L,R}$ in terms of the
VLL mass eigenvalues,
\begin{equation}
\begin{aligned}
\lambda_L &= -\,\frac{\sqrt{2}}{v \,c_\beta}\,\left[c_R s_L M_1\,-\,c_L s_R 
M_2\right] \,,\\
\lambda_R &= -\,\frac{\sqrt{2}}{v \,c_\beta}\,\left[c_L s_R M_1\,-\,c_R s_L 
M_2\right]\,.
\end{aligned}
\end{equation}
We see that high $\tan\beta$ will increase the magnitude of these couplings, as will higher
values of $M_1$, $M_2$, unless some fine-tuning with the angles $\theta_{L,R}$ occurs.

\begin{figure}[t!]
	\centering
		\includegraphics[width=0.47\textwidth]{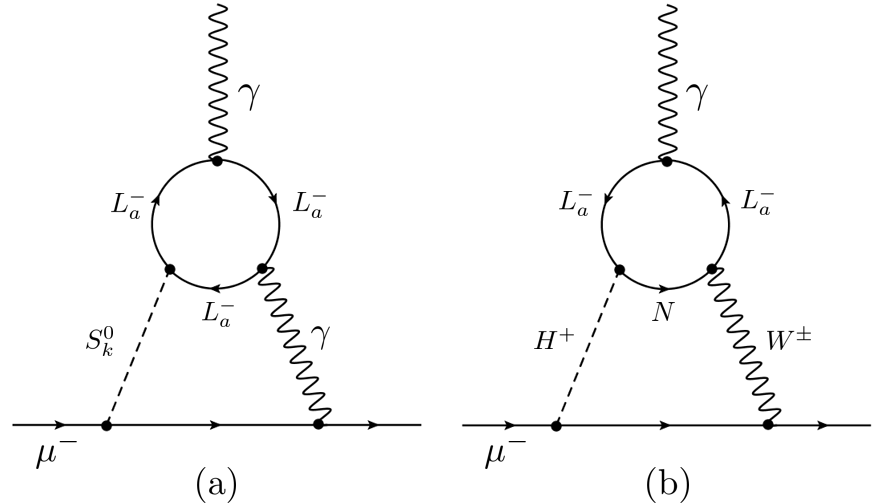}
\caption{BZ contributions to $\Delta a_\mu$ involving the new charged and neutral leptons $L_a^-$ and $N$, respectively. }
	\label{fig:fig2}
\end{figure}

In the absence of VLL mixing with the muon, there are no new one-loop contributions to $\Delta a_\mu$ besides those already present in the 2HDM. Still, BZ diagrams involving the new charged and neutral leptons must be taken into account (see Fig.~\ref{fig:fig2}).
We have computed both contributions and, for diagram (a), our result agrees with that of Ref.~\cite{Ilisie:2015tra}, doing the appropriate replacements according to the interactions shown in \eqref{eq:LSHVLL}. Namely\footnote{Our result for this diagram also agrees with that of Ref.~\cite{Chun:2020uzw}.},
\begin{align}
  \Delta a_\mu^{\rm (a)} = &\dfrac{\alpha \, m_\mu^2}{4 \pi^3 v^2} \sum_{k,a} Q_a^2 \Biggl[ \text{Re}\left( y_{aa}^k \right) \text{Re}\left( y_{\mu}^k \right) \mathcal{F}^{(1)} \left(      \dfrac{M_a^2}{M_k^2}\right)\Biggr.\nonumber\\
    &+ \, \Biggl.\text{Im}\left( y_{aa}^k \right) \text{Im}\left( y_{\mu}^k \right) \mathcal{F}^{(2)} \left(\dfrac{M_a^2}{M_k^2}\right)\Biggr] \, ,
\end{align}
with
\begin{equation}
    \mathcal{F}^{(1,2)}(r)=\dfrac{r}{2}\int_0^1 dx \dfrac{\mathcal{N}^{(1,2)}}{r-x(1-x)} \text{ln} \left[ \dfrac{r}{x(1-x)} \right] \, ,
\end{equation}
and $\mathcal{N}^{(1)}=2x(1-x)-1$ and $\mathcal{N}^{(2)}=1$.
Regarding diagram (b), caution must be taken since, contrarily to what happens in the 2HDM, charged-current interactions are now left- and right-handed -- see Eq.~\eqref{eq:LCC}. 
We have computed the contribution from this diagram, the result being:
\begin{align}
  \Delta a_\mu^{\rm (b)} =& \dfrac{\alpha \, m_\mu^2}{32 \pi^3 s_W^2 v^2 \left( M^2_{H^+}-M_W^2\right)} \sum_{a} \int_0^1 dx \, Q_a (1-x) \nonumber \\
  & \times \left[ \mathcal{G}\left(\dfrac{M_N^2}{M^2_{H^+}},\dfrac{M_a^2}{M^2_{H^+}} \right) - \mathcal{G}\left(\dfrac{M_N^2}{M^2_{W}},\dfrac{M_a^2}{M^2_{W}} \right) \right]\nonumber\\
  & \times \bigl( \omega_L + \omega_R \bigr) \, ,
\end{align}
where the function $\mathcal{G}$ is
\begin{equation}
    \mathcal{G}(r_1,r_2)=\dfrac{\text{ln}\left[ \dfrac{r_1 x + r_2 (1-x)}{x(1-x)}\right]}{x(1-x)-r_1 x - r_2 (1-x)} \, .
\end{equation}
The term proportional to $\omega_{L(R)}$ takes the left (right)-handed component of the charged-current interactions with the new leptons, such that:
\begin{equation}
\begin{aligned}
\omega_L&= \text{Re}\left[ (U_L)_{1a}^{*} \, \xi_a M_a (U_R)_{2a} \, \xi_\mu^{*} \right]M_a \left[ (1-x) - (1-x)^2 \right]\\
&- \text{Re}\left[ (U_L)_{1a}^{*} \, \xi_N M_N (U_L)_{2a} \, \xi_\mu^{*} \right]M_N x (1 + x) \,,\\
\omega_R&= - \text{Re}\left[ (U_R)_{1a}^{*} \, \xi_N M_N (U_L)_{2a} \, \xi_\mu^{*} \right]M_a \left[ (1-x) + (1-x)^2 \right] \\
&+ \text{Re}\left[ (U_R)_{1a}^{*} \, \xi_a M_a (U_R)_{2a} \, \xi_\mu^{*} \right] M_N \, x(1-x)\,.
\end{aligned}
\end{equation}
The first term is analogous to that for quarks presented in Ref.~\cite{Ilisie:2015tra}, while the second has the same structure as the general terms obtained in Ref.~\cite{Yang:2009zzh}.

While fitting the experimental values of $(g - 2)_\mu$ within the context of a 2HDM, we must make sure 
that all experimental and theoretical constraints of that model are satisfied. We have already mentioned $B$-physics constraints -- particularly relevant for the type-II model, wherein $b\rightarrow s \gamma$ 
results force the charged Higgs mass to be very high -- and bounded from below and unitarity bounds,
which limit the values of the scalar potential's quartic couplings (see ~\cite{Branco:2011iw} for 
the explicit expressions). Equally important is to consider electroweak precision constraints, in the form
of the Peskin-Takeuchi parameters $S$, $T$ and $U$~\cite{Peskin:1990zt}. Their explicit expression for the 2HDM
can be found in numerous references (see for instance eqs. (12) and (13) of~\cite{Kanemura:2011sj}),
but if one  considers vector-like leptons present as well, their contributions to the $T$ parameter 
must also be taken into account (see Eq. (4.5) of~\cite{Chun:2020uzw}) -- the upshot of their inclusion
in a 2HDM fit is that it usually is always possible, for VLL masses $M_1$ and $M_2$ below $\sim$ 500 GeV, 
to find a vector-like neutrino mass $m_N$ to satisfy the current constraints on $T$. Indeed, these electroweak precision constraints, as is usual in the vanilla 2HDM, tend to reduce the splittings between the extra (scalar and VLL) masses in the theory.

With or without VLLs, LHC precision constraints on the properties of the 125 GeV $h$ scalar are,
with the exception of the diphoton branching ratio, trivially satisfied within the alignment limit, since $h$'s tree-level couplings become identical to the SM Higgs when 
$\cos(\beta - \alpha) = 0$. The presence of a charged scalars and two charged VLLs, however,
contribute to the diphoton decay width. The decay amplitude $\mathcal{A}$ for $h\rightarrow\gamma\gamma$
includes therefore a contribution identical to the SM, and another from the charged and 
VLL sector, given by
\begin{equation}
\mathcal{A} = \mathcal{A}_{\rm SM} - \frac{\Lambda}{2 m^2_{H^+}}A^H_{0}(\tau_+) + y^h_{11} A^H_{1/2}(\tau_1) + 
y^h_{22} A^H_{1/2}(\tau_2) \,,
\label{eq:hgg}
\end{equation}
where $\tau_X = m_h^2/(4 m^2_X)$, $\Lambda = 3 m^2_h - 2 m^2_{H^+} - 4 m^2_{12}/s_{2\beta}$ and
$A^H_0$, $A^H_{1/2}$ are the well-known scalar and fermionic form factors of the Higgs (see for 
instance~\cite{Djouadi:2005gj}). Depending on the charged and VLL masses and couplings, their 
contributions  to $h\rightarrow\gamma\gamma$ can actually cancel each other, and will be less 
important for higher masses.
\begin{figure}[h!]
	\centering
	\begin{tabular}{l}
		\includegraphics[width=0.425\textwidth]{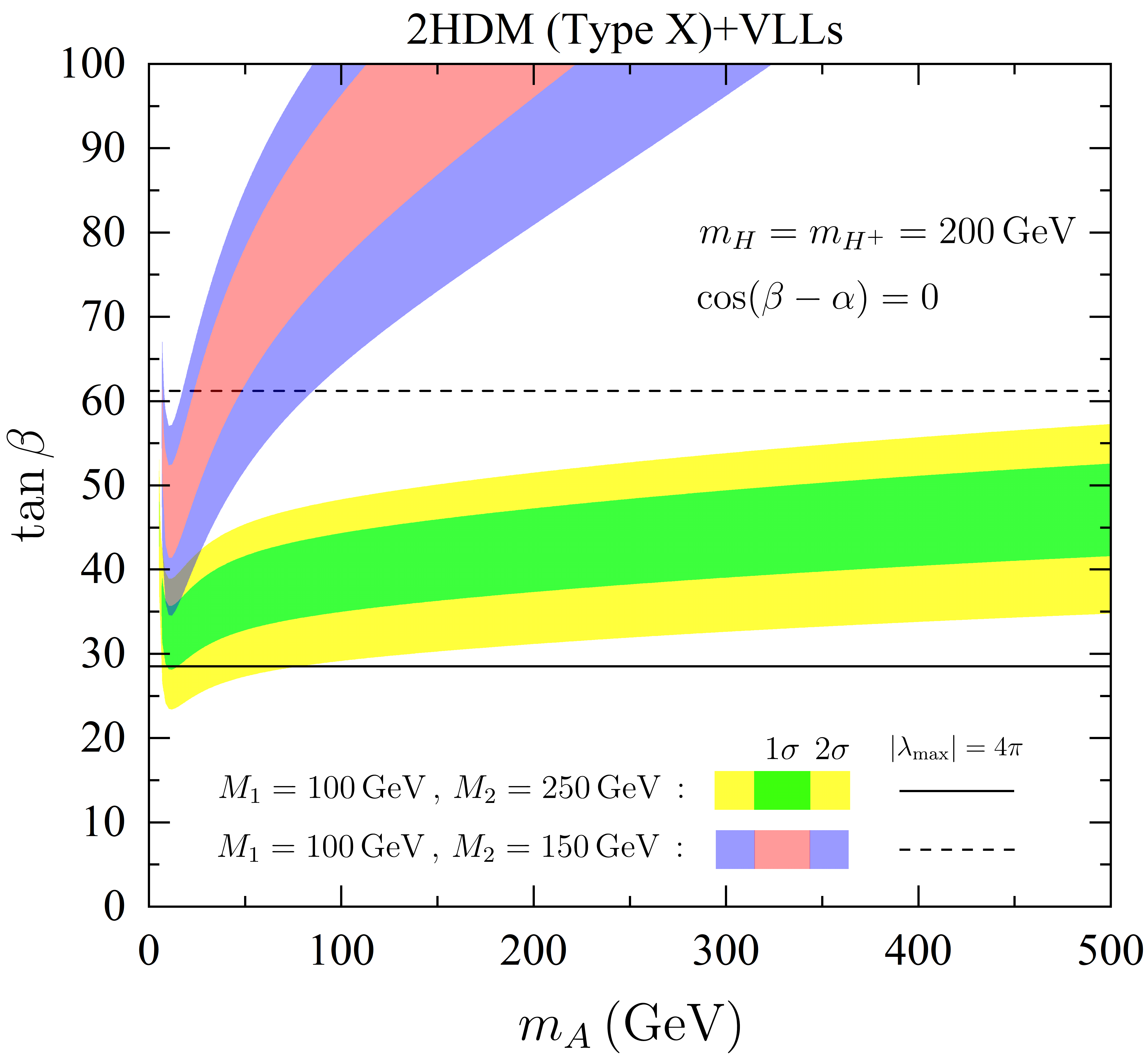}\\
	\includegraphics[width=0.425\textwidth]{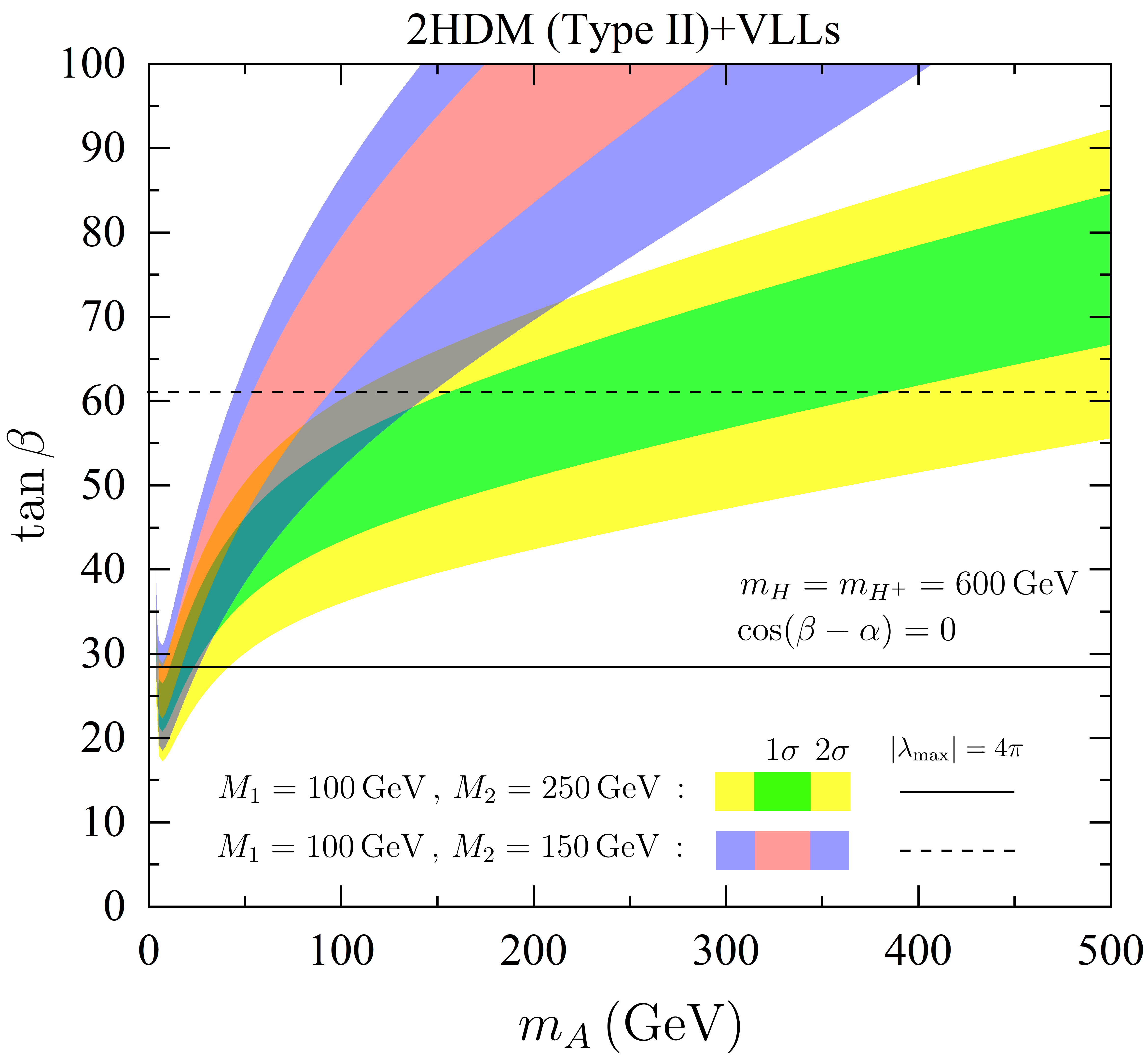}\\
	\includegraphics[width=0.425\textwidth]{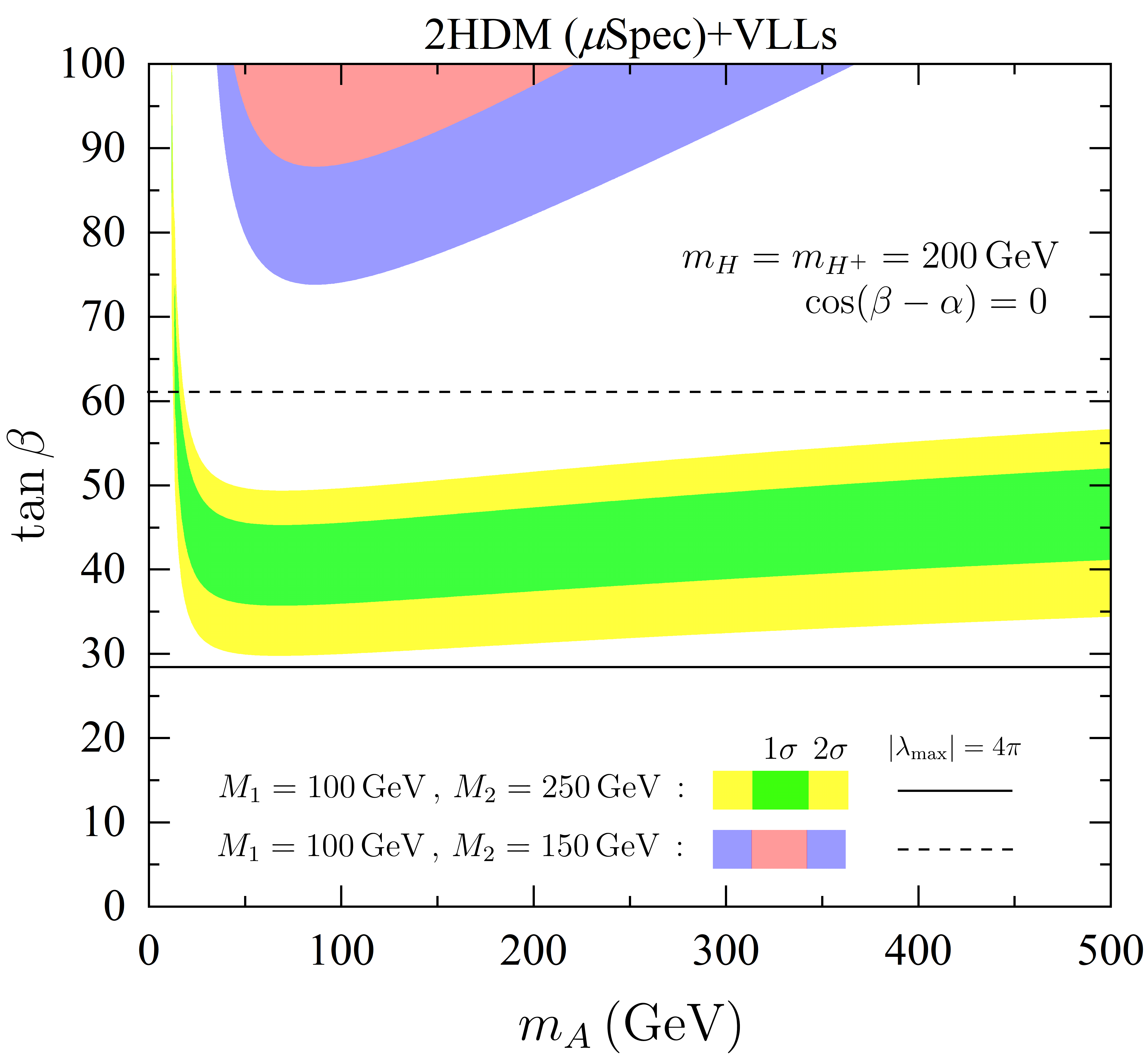}
	\end{tabular}
\caption{$(m_A,\tan\beta)$ allowed regions for the type-X, type-II and $\mu$Spec 2HDM with VLLs (from top to bottom panel),taking the benchmarks B1 and B2 given in \eqref{B1B2}. The values $s_L = 0.5, s_R = 0.4$ are chosen. 
Above the solid (dashed) horizontal line, $\lambda_{\rm max}\equiv {\rm max}\{|\lambda_L|,|\lambda_R|\}>4\pi$ 
for B$_1$ (B$_2$). See text for more details.} 
\label{fig3}
\end{figure} 

To illustrate the main features of the VLL models we are interested in, we focus on two benchmark cases (B$_1$ and B$_2$) within the type-II, type-X and $\mu$Spec 2HDM with heavy charged-lepton masses:
\begin{align}
    {\rm B}_1:M_2=2.5M_1=250\,{\rm GeV}\,,\nonumber\\
    {\rm B}_2:M_2=1.5M_1=150\,{\rm GeV}\,,
    \label{B1B2}
\end{align}
taking $s_L(s_R)=0.5(0.4)$. The neutral-lepton mass $M_N$ turns out to be fixed since $M_N=|m_L|=|M_1c_Lc_R+M_2s_Ls_R|$, yielding $M_N\simeq 129\,(109)$~GeV for B$_1$ (B$_2$). The scalar masses $m_{H^+}$ and $m_{H}$ are the same as those considered in Fig.~\ref{fig1} (for $\mu$Spec we adopt the same setting as for type X) and, again, $\cos(\beta-\alpha)=0$.\footnote{The impact on $\Delta a_\mu$ of considering $\cos(\beta-\alpha)\neq 0$ within the experimentally-allowed limits is marginal. For instance, taking $\cos(\beta-\alpha) = 0.05$ the change in the results is almost imperceptible.} In Fig.~\ref{fig3} we show the $(m_A,\tan\beta)$ allowed regions in green and yellow (blue and red) for B$_1$ (B$_2$) at 1 and $2\sigma$, respectively. Above the solid (dashed) horizontal line, $\lambda_{\rm max}\equiv {\rm max}\{|\lambda_L|,|\lambda_R|\}>4\pi$ for B$_1$ (B$_2$). 

The results show that in all cases the ranges of $m_A$ can be substantially enlarged with respect to Fig.~\ref{fig1}, while simultaneously shifting $\tan\beta$ to lower values. This effect is more pronounced for B$_1$ since $M_2/M_1$ is larger than for B$_2$, and so are the $L_2L_2 S^0_k$ couplings -- see Eqs.~\eqref{eq:xiVLL} and \eqref{eq:xpm}. The main drawback of this scenario is that the larger VLL coupling with the Higgs doublet $\Phi_1$ is required to be close or above the perturbative limit, i.e. $\lambda_{\rm max}=4\pi$. This can be easily understood noting that the coupling modifiers $\xi_{a,N}$ in Eq.~\eqref{eq:xiVLL} scale as $ \lambda v \cos\beta/(\sqrt{2}M)$ (with respect to the pure 2HDM case) where $\lambda$ and $M$ stand for a generic VLL coupling and mass, respectively. Requiring that, at least, $\xi_{a,N}\simeq \tan\beta$ implies $\lambda v \cos\beta/(\sqrt{2}M)\simeq 1$. Thus, for $M\gtrsim 100$~GeV and large $\tan\beta$, $\lambda$ must be large to avoid spoiling the $\tan\beta$ enhancement required to explain the $\Delta a_\mu$ discrepancy in 2HDMs. For the $\mu$Spec model, the effect of adding VLLs allows to drastically lower $\tan\beta\gtrsim \mathcal{O}(1000)$ down to $\tan\beta\sim 40-50$ for $m_A \gtrsim 200$~GeV.

From an analysis of Fig.~\ref{fig3}, we conclude that a strict requirement of perturbativity on the 
$\lambda_{L,R}$ couplings would provoke a drastic curtailment of the available parameter space -- for the benchmarks
presented, the $\mu$Spec model would have no available parameter space left, since the 1/2-$\sigma$ bands shown
there always lie {\em above} the corresponding $\lambda_{\rm max} = 4\pi$ lines. For the type II and X models the only
surviving region is the low pseudoscalar mass one ($m_A$ smaller than roughly, respectively, 150 and 100 GeV for the
second benchmark shown), albeit with larger masses than those obtained without VLLs. Nonetheless, the low masses for
pseudoscalars obtained for the type II model cannot be accomodted in that model, given that the $b\rightarrow s 
\gamma$ constraints imposed a lower bound on the charged mass of, generously, 600 GeV. A possible way to enlarge the allowed region would be to increase the number $N$ of VLL generations, since {\em na\"{\i}vely} one would expect that this limit would be improved by $\lambda_{L,R}/N$.

Since the benchmark analysis shows type X is favoured for agreement with the muon anomaly, we have performed 
a general parameter space scan of the 2HDM for that model, considering extra scalars 
and VLLs (charged and neutral) to have masses ranging from 100 to 1000 GeV, as well as all 
possible values for the angles $\theta_L$ and $\theta_R$. We only accepted combinations of 
parameters for which the model satisfies unitarity; boundedness from below; electroweak precision 
constraints; $b\rightarrow s \gamma$ constraints; and predicts a Higgs diphoton branching ratio within 10\% of the SM values. Finally,
since extra scalars are required to explain the $(g-2)_\mu$ anomaly, we need to verify whether current LHC searches would not have discovered those particles already. The $B$-decay constraints we mentioned take care of the charged Higgs; because we are working in the alignment limit, searches for the heavier CP-even scalar $H$ in $ZZ$ or $WW$ are automatically satisfied; indeed, the channel which we need to worry the most would be searches for $H$ and specially $A$ decaying into tau pairs, having been produced via gluon fusion. Tau decays can be $\tan\beta$-enhanced
and produce strong signal rates already excluded by LHC collaborations~\cite{Aaboud:2017sjh}. 
We verified that, after all other constraints had been taken into account, our scan included only a small fraction of points above current experimental limits on ditau searches for extra scalars,
and excluded those points from our $(g - 2)_\mu$ calculations. The result of that scan show that the type X behaviour shown in Fig.~\ref{fig3} can be reproduced by a wealth of different combinations of parameters; the values of allowed $\tan\beta$ are even more general, with, for instance, $\tan\beta \simeq 25$ permitted for $m_A = 400$ GeV. However, and also confirming the previous conclusions, though unitarity and perturbativity can be satisfied within
the scalar sector, agreement with $(g - 2)_\mu$ requires large VLL yukawas -- in particular, though $\lambda_L$ 
can be found small, $|\lambda_R|$ is always found to be above $\sim$ 20, and thus non-perturbative.

In \cite{Chun:2020uzw} the authors work with a small mass difference between the charged-vector-like masses and do not find any problems regarding perturbativity of the VLL yukawas, which appears to contradict our results. Without taking into account all the above constraints, and working on that same region, we were able to reach that same result. However, an exhaustive scan as the one we have performed completely excludes such a region of the parameter space and a small mass difference between $M_1$ and $M_2$ is not able to reproduce the $(g-2)_\mu$ anomaly. Finally, a word 
on limits on $\tan\beta$ stemming from tau and $Z$ decays, as considered in~\cite{Chun:2020uzw} -- though in that
work mixing between VLLs and electrons is allowed and therefore a direct comparison is not practical, their Fig. 4 
seems to show ample parameter space available for the case of a scalar spectrum without too large mass splittings,
as is the case of our Fig.~\ref{fig3} or the result of eletroweak precision constraints in our general Type X
scan. However, the point remains that the non-perturbativity found for the couplings $\lambda_{L,R}$ is a
stronger limitation on the validity of this approach to fitting $g - 2$ that the bounds stemming from tau and 
$Z$ decays.

\section{Conclusions} 

We have studied the theory/experiment discrepancy of the muon's anomalous magnetic moment in the 2HDM. Without vector-like leptons, the type-I and type-Y models cannot explain the discrepancy and the type II requires light pseudoscalars that are in conflict with perturbativity, unitarity and electroweak precision constraints. The type-X model can accommodate the discrepancy but require large values of $\tan\beta$ and also very light pseudoscalars. After revisiting some 2HDM results in light of the new result reported by the Muon g-2 collaboration at Fermilab, we considered adding vector-like leptons to the type-II, type-X and muon-specific 2HDMs. We did not include any mixing between the vector-like leptons and the muon. We have shown that the parameter space is substantially widened and much larger values of the pseudoscalar mass are allowed. $\tan\beta$ remains fairly large in all of these models. One problem of the VLL scenario analysed in this work is that the VLL Yukawa couplings to the scalar doublet $\Phi_1$ must be close to or above the perturbation theory limit. This can be alleviated by considering $N$ families of vector-like leptons instead of only one (very roughly, this would reduce the maximum values of these couplings by a factor of $1/N$). Alternatively, one can include VLL-muon mixing. This has been done in Refs~\cite{Dermisek:2013gta,Dermisek:2021ajd} in some detail, where it was also shown that the two-loop Barr-Zee diagrams are substantially less important than one-loop contributions and that smaller values of $\tan\beta$ can be accommodated. Finally, one can relax our assumption that a symmetry eliminates tree-level FCNC and consider the A2HDM mentioned in the introduction; this would add additional parameters and might alleviate the problem of perturbativity.\\

{\bf Acknowledgements:} This work is supported by Funda\c{c}{\~a}o para a Ci{\^e}ncia e a Tecnologia (FCT, Portugal) through the projects UIDB/00777/2020, UIDP/00777/2020, UIDB/00618/2020, CERN/FIS-PAR/0004/2019, CERN/FISPAR/0014/2019, PTDC/FIS-PAR/29436/2017 and by HARMONIA project's contract UMO-2015/18/M/ST2/00518. The work of B.L.G. is supported by the FCT grant SFRH/BD/139165/2018. The work of M.S. was supported by the National Science Foundation under grant PHY-1819575

\end{document}